\documentclass[twocolumn,superscriptaddress,amsmath,amssymb,aps,longbibliography,prl]{revtex4}
\usepackage[colorlinks=true,urlcolor=blue,citecolor=blue,linkcolor=blue]{hyperref}
\usepackage{txfonts}
\usepackage{amsfonts}
\usepackage{amsmath}
\usepackage{graphicx}
\usepackage{diagbox}
\usepackage{amssymb}
\usepackage{bbm}
\usepackage{float}
\usepackage{subfigure}
\usepackage{url}
\usepackage{hyperref}
\usepackage{multirow}
\usepackage{bm}
\usepackage{booktabs}
\usepackage{array}
\usepackage{color}
\usepackage{algorithm}
\usepackage{algorithmic}

\newcommand{\PreserveBackslash}[1]{\let\temp=\\#1\let\\=\temp}
\usepackage{makecell}
\newcolumntype{C}[1]{>{\PreserveBackslash\centering}p{#1}}
\newcolumntype{R}[1]{>{\PreserveBackslash\raggedleft}p{#1}}
\newcolumntype{L}[1]{>{\PreserveBackslash\raggedright}p{#1}}

\usepackage{xcolor}
\definecolor{red}{rgb}{0.7, 0.1, 0.1}
\definecolor{green}{rgb}{0.0, 0.5, 0.0}

\begin{document}

\title{Quantum Anharmonic Effects in Hydrogen-Bond Symmetrization of High-Pressure Ice}

\author{Qi Zhang}
\affiliation{Beijing National Laboratory for Condensed Matter Physics and Institute of Physics, Chinese Academy of Sciences, Beijing 100190, China}

\author{Lei Wang}
\email{wanglei@iphy.ac.cn}
\affiliation{Beijing National Laboratory for Condensed Matter Physics and Institute of Physics, 
Chinese Academy of Sciences, Beijing 100190, China}

\date{\today}
	
\begin{abstract}
    The nuclear quantum effects of hydrogen play a significant role in determining the phase stability of water ice. 
    Hydrogen-bond symmetrization occurs as hydrogen atoms tunnel in a double-well potential, ultimately occupying the midpoint between oxygen atoms and transforming ice VIII into ice X under high pressure.
    Quantum fluctuations lower this transition from classical predictions of over $100~\text{GPa}$ to $60~\text{GPa}$.
    We reveal that the Perdew-Burke-Ernzerhof functional underestimates the hydrogen double-well barrier, thus resulting in a transition pressure over $10~\text{GPa}$ lower than the strongly constrained and appropriately normed functional, which is validated against quantum Monte Carlo calculations.
    Nuclear quantum anharmonicity, treated via neural canonical transformation (NCT),  reveals that this transition pressure is temperature-independent and observes a $2~\text{GPa}$ reduction when comparing the non-Gaussian flow models based wavefunction compared to the self-consistent harmonic approximation.
    Despite increasing pressure typically shortens chemical bonds and hardens phonon modes, NCT calculations reveal that hydrogen bond softens hydrogen-oxygen stretching in ice VIII upon pressure.

\end{abstract}
\maketitle

\begin{figure}[htbp]

	\centering
    \hspace{-4.0mm}
	\includegraphics[width=0.50\textwidth]{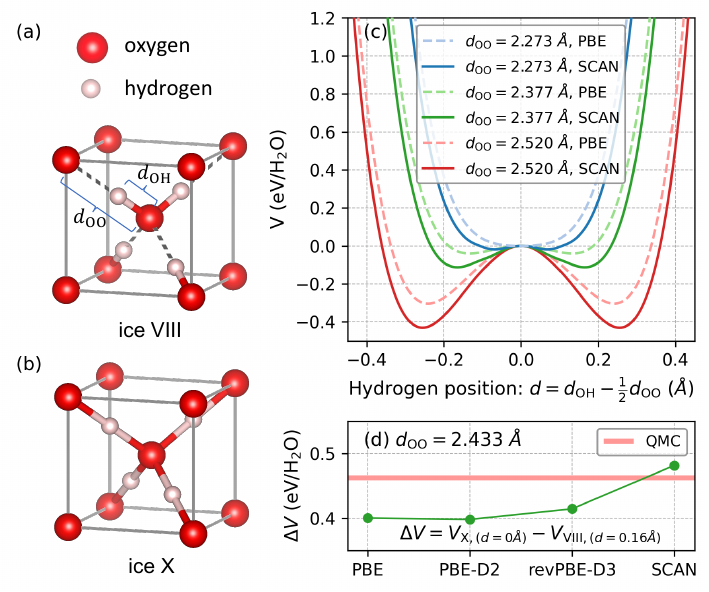}
	\caption{
    (a) The sketch of ice VIII, visualized using VESTA~\cite{momma2011vesta}.
    The illustration shows only part of the VIII unit cell, which contains oppositely polarized configurations for net-zero polarization.
    (b) Ice X.
    (c) The potential energy surface is calculated using various density functional theory functionals. 
    All hydrogen atoms are moved along the O-O direction, with their positions defined in (a) while the oxygen atoms remain fixed. 
    (d) The energy difference of ice X and ice VIII computed with various density functionals are compared with QMC results. 
    }
	\label{fig1}
\end{figure}

\textit{Introduction}.--
Water is one of the most abundant hydrogen-containing compounds in nature.
In water ice, 
each molecule can form up to four hydrogen bonds with neighboring molecules, creating a unique hydrogen-bonded network that gives rise to a remarkable diversity of crystal structures~\cite{loerting2020open}.
At ambient conditions, each hydrogen atom forms a covalent bond ($\text{O}-\text{H}$) with one oxygen atom and a hydrogen bond ($\text{O}\cdots \text{H}$) with another.
However, increasing pressure reduces molecular spacing and distorts bond angles, leading to a series of phase transitions. 
Above $2~\text{GPa}$, oxygen atoms form a simple body-centered cubic structure, known as ice VII~\cite{bridgman1937phase, kamb1965overlap}. 
It transforms into the proton-ordered ice VIII [Fig.~\ref{fig1}~(a)]~\cite{pruzan1990raman,pruzan1992determination,goncharov1999raman} at low temperatures, exhibiting unit cell slightly elongation along the $c$-axis~\cite{besson1994variation,pruzan2003phase,fukui2022equation}. 
Extensive studies have shown that at pressures over $60~\text{GPa}$,
water's molecular structure breaks down, transforming into an atomic solid, labeled as ice X [Fig.~\ref{fig1}~(b)],
where hydrogen atoms occupy in the center of two oxygen atoms, forming a symmetric hydrogen configuration~\cite{schweizer1984high,hirsch1986effect,lee1992ab,lee1993ab,aoki1996infrared,goncharov1996compression,struzhkin1997cascading,song1999infrared,benoit1998tunnelling,bernasconi1998abinitio,loubeyre1999modulated,caracas2008dynamical,lin2011correlated,meier2018observation,li2024phase,cherubini2024quantum,komatsu2024hydrogen,monacelli2025hydrogen}.
The symmetrization of hydrogen bonds is a fundamental phenomenon, providing insights into the behavior of hydrogen-bonded systems under extreme conditions.

Under the Born-Oppenheimer approximation, the transition mechanism from ice VII/VIII to X is well understood~\cite{bernasconi1998abinitio,komatsu2024hydrogen,kuwahata2024nuclear}, being a localized process independent of thermodynamics.
This transition is driven by the evolution of the double-well potential energy surface (PES) experienced by hydrogen atoms situated between two oxygen atoms, as illustrated in Fig.~\ref{fig1}~(c).
At low pressures, hydrogen atoms are confined to one of the wells by a barrier on the order of several electron volts ($\text{eV}$).
By $60~\text{GPa}$, this barrier decreases to the scale of a few hundred $\text{meV}$~\cite{pruzan2003phase,caracas2008dynamical,lin2011correlated,bronstein2014quantum}, enabling quantum tunneling of hydrogen atoms and facilitating the formation of ice X.
Above $120~\text{GPa}$, the PES transforms into a single well, and the hydrogen atom becomes fully localized at the midpoint, resulting in a highly symmetrical hydrogen structure understood well in classical picture~\cite{kuwahata2024nuclear, cherubini2024quantum}.

\begin{table*}[htbp]
    \centering
    \caption{Transition pressures for the VII/VIII-X phase transition as obtained using various nuclear and electronic methods. 
    $^\text{a}$ Linear combination of two displaced Gaussian wavefunctions.
   }
    \label{tab:1}
    \setlength{\tabcolsep}{5pt}
    \begin{tabular}{C{1.9cm}C{1.0cm}C{5.1cm}C{4.5cm}C{3.5cm}}
        \addlinespace[3pt]
        \hline\hline
        \addlinespace[3pt]
        \textbf{Reference} & 
        \textbf{Year} &
        \textbf{Nuclear method}  &
        \textbf{Electronic method} & 
        \textbf{Transition pressure} (GPa) \\
        \addlinespace[3pt]
        \hline
        \addlinespace[3pt]

        \cite{schweizer1984high}
        & 1984
        & Non-Gaussian$^\text{a}$  (variational)
        & Empirical (Morse potential)
        & 45 
        \\

        \cite{lee1992ab}
        & 1992  
        & Gaussian (variational)
        & GGA (Becke 1988)
        & 49 
        \\

        \cite{benoit1998tunnelling} 
        & 1998
        & Non-Gaussian (PIMD)
        & GGA (Becke 1988)
        & 72 
        \\

        \cite{kang2013quantum}
        & 2013
        & Non-Gaussian (PIMD)
        & GGA (PBE-D2)
        & 61.6
        \\

        \cite{bronstein2014quantum}
        & 2014
        & Non-Gaussian (Quantum thermal bath)
        & GGA (PBE)
        & 65
        \\

        \cite{kuwahata2024nuclear} 
        & 2024
        & Non-Gaussian (PIMD)
        & GGA (revPBE-D3)
        & 61 
        \\

        \cite{cherubini2024quantum} 
        & 2024
        & Non-Gaussian (PIMD)
        & GGA (PBE)
        & 43
        \\

        \cite{cherubini2024quantum}
        & 2024
        & Gaussian (SSCHA)
        & GGA (PBE)
        & 51
        \\

        \textbf{Present work}
        & \textbf{2025}
        & \textbf{Non-Gaussian (NCT)}
        & \textbf{meta-GGA (SCAN)}
        & \textbf{60.6}
        \\
        \addlinespace[3pt]
        \hline\hline
    \end{tabular}
\end{table*}

Spectroscopic experiments have identified the transition to ice X occurring at pressures between $58$ and $62~\text{GPa}$~\cite{aoki1996infrared,goncharov1996compression,struzhkin1997cascading,song1999infrared}, but theoretical calculations exhibit considerable discrepancies~\cite{schweizer1984high,lee1992ab,benoit1998tunnelling,kang2013quantum,bronstein2014quantum,kuwahata2024nuclear,cherubini2024quantum}, as shown in Tab.~\ref{tab:1}.
Early studies have found the limitations of the local density approximation (LDA) for hydrogen bonds,
with generalized gradient approximation (GGA) more accurate~\cite{lee1992ab,lee1993ab}.
Subsequent work~\cite{benoit1998tunnelling} using the path-integral molecular dynamics (PIMD) technique with GGA and predicted $P_c = 72~\text{GPa}$.
Recently, Ref.~\cite{cherubini2024quantum} employed the stochastic self-consistent harmonic approximation (SSCHA) with a machine learning potential from the Perdew-Burke-Ernzerhof (PBE)~\cite{perdew1996generalized} functional, estimating $P_c = 51~\text{GPa}$.
However, their PIMD simulations surprisingly yielded a much lower value of $P_c = 43~\text{GPa}$, further deviating from experimental results.

These observations highlight the necessity of accurately determining nuclear quantum effects, anharmonic dynamics, and electronic structure calculations in high-pressure ice. 
We compare various density functional theory (DFT) functionals, along with quantum Monte Carlo (QMC)~\cite{li2022deepsolid} calculations, to assess their influence.
To capture quantum anharmonicity and thermal effects, we implement the Neural Canonical Transformation (NCT) approach~\cite{xie2022abinitio, xie2023mstar, zhang2024neural, zhang2024lithium}.
This method combines a normalizing flow model~\cite{dinh2014nice, dinh2016density, papamakarios2019neural, papamakarios2021normalizing, saleh2023computing}  to describe phonon excited-state wavefunctions with a probability model for energy level occupations, enabling joint optimization of free energy, as detailed in our previous work~\cite{zhang2024neural, zhang2024lithium} and Supplemental Material (SM)~\cite{supp}.
NCT also directly provides the phonon density of states beyond the harmonic approximation.
The code is open-source and publicly available~\cite{mycode}.

\textit{Electronic structure calculations and potential energy surface}.--
Accurate determination of the PES is critical for reliable studies of high-pressure water ice.
The widely used PBE functional often underestimates the double-well barrier height~\cite{verdi2023quantum}.
Previous studies have highlighted the importance of van der Waals dispersion interactions in crystalline ice phases~\cite{santra2011hydrogen,murray2012dispersion}. 
However, Ref.~\cite{bronstein2014quantum} shows that van der Waals contributions are negligible in high-pressure ice.
The strongly constrained and appropriately normed semilocal density (SCAN) functional~\cite{sun2015strongly}, a meta-GGA, offers more accurate exchange-correlation energy estimates.
It has demonstrated diffusion Monte Carlo (DMC) level accuracy for ice VIII and shows excellent agreement with experimental data, outperforming PBE~\cite{sun2015scan}.
Similarly, Ref.~\cite{della2022dmc} reports that SCAN provides better accuracy than PBE for both VII and VIII phases when compared to DMC benchmarks. 
However, neither study investigates the performance of SCAN for ice X.


To assess the performance of these exchange-correlation functionals under high pressures, we fixed the oxygen atoms and moved the hydrogen atoms within a periodic box containing $2~\text{H}_2\text{O}$, analyzing potential energy changes.
Due to the high cost of QMC calculations, the simulations were restricted to the $\Gamma$ point.
In Fig.~\ref{fig1}~(d), the oxygen-oxygen distance, representing the intermolecular spacing, is set to $d_{\text{OO}} = 2.433~\AA$, corresponding to the pressure near $50~\text{GPa}$. 
The value $d = 0~\AA$ represents the midpoint of the double-well potential, 
representing the symmetric hydrogen structure of ice X,
while $d = 0.16~\AA$ indicates the asymmetric ice VIII structure.
The energy differences between PBE, PBE-D2, and revPBE-D3 are within $15~\text{meV}/\text{H}_2\text{O}$, suggesting that dispersion corrections are not significant for high-pressure ice, which is consistent with previous findings~\cite{bronstein2014quantum}.
SCAN closely matches QMC results, with a difference of $20~\text{meV}$ per $\text{H}_2\text{O}$, while PBE shows a discrepancy of $60~\text{meV}$.
These results underscore the accuracy of SCAN for water ice at extreme pressures.

Due to strong finite-size effects in the electronic calculations at the $\Gamma$ point, we analyzed the double-well potential more precisely by calculating the PES in a $16~\text{H}_2\text{O}$ cell with a $6\times6\times6$ $k$-point mesh, as shown in Fig.~\ref{fig1}~(c).
At equivalent densities, SCAN consistently predicts higher barriers than PBE. 
While the oxygen-oxygen distance $d_{\text{OO}}=2.273~\AA$, corresponding to a pressure of approximately $30~\text{GPa}$, PBE underestimates the barrier height by over $120~\text{meV}$ relative to SCAN.
However, hydrogen tunneling is negligible under this condition and does not affect the stability of the VIII structure.
At higher densities ($d_{\text{OO}}=2.520~\AA$, $120~\text{GPa}$) PBE predicts that the double-well barrier disappears, while SCAN indicates a $40~\text{meV}$ barrier.
This discrepancy significantly underestimates the ice VIII-X transition pressure when using PBE.

To quantify the impact of different functionals on the transition pressure, our calculations show that the SCAN functional predicts a barrier height of $140~\text{meV}$ at $60~\text{GPa}$, which allows hydrogen tunneling to occur.
In contrast, PBE requires only $47~\text{GPa}$ to produce the same barrier height, as shown in~\cite{supp}.
Based on these results, we roughly estimate that the VIII-X transition pressure predicted by PBE is more than $10~\text{GPa}$ compared to SCAN. 
This aligns with Ref.~\cite{cherubini2024quantum}, where PIMD calculations with PBE report a transition pressure of $43\text{GPa}$.

Building on these insights, we adopt the SCAN functional as the optimal choice in the following calculation. 
To balance accuracy and efficiency, we employed the high-order equivariant message-passing neural networks, specifically a machine learning potential known as the multi-atomic cluster expansion (MACE)~\cite{batatia2022mace, batatia2022design}, to fit the energy and force data.
The MACE model achieved excellent performance on high-pressure ice, with a root mean square error of $0.1~\text{meV}/\text{atom}$ for energy and $3~\text{meV}/\AA$ for forces on the test set~\cite{supp}.

\begin{figure}[htbp]
	\centering
    \hspace{-4.0mm}
	\includegraphics[width=0.50\textwidth]{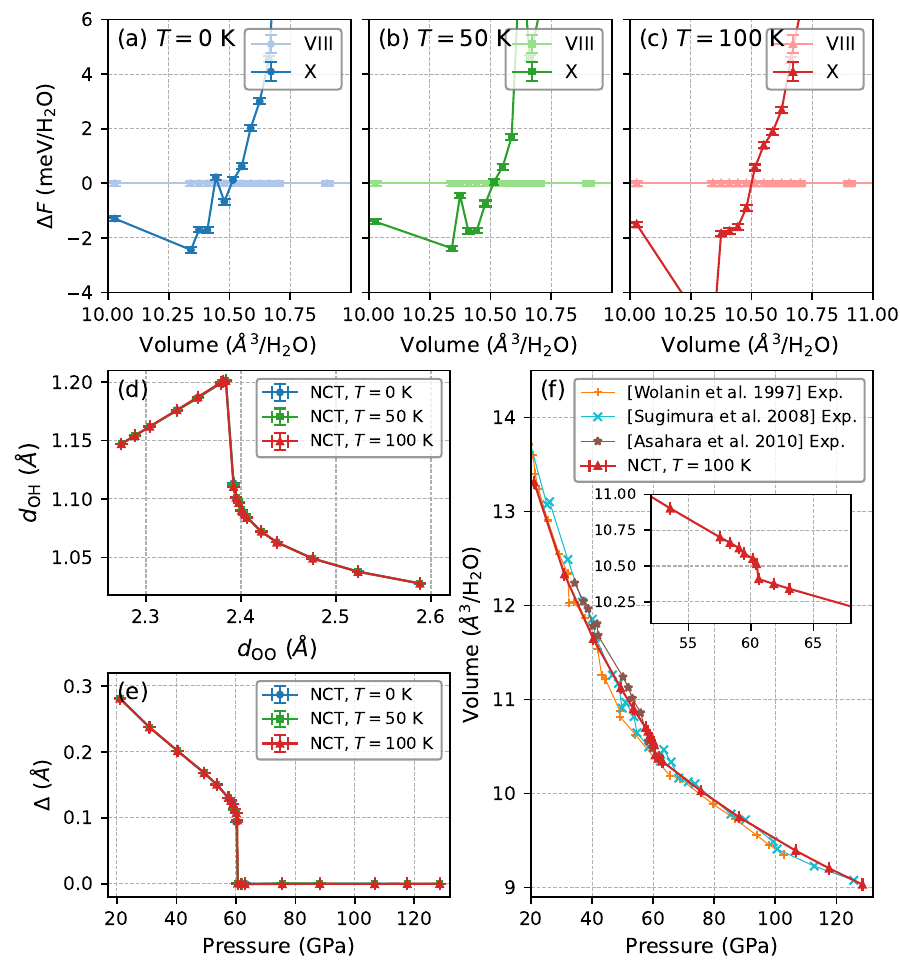}
	\caption{
        Numerical results of the NCT method for high-pressure ice.
        Helmholtz free energy differences $\Delta F$ between ice VIII and X as a function of volume at (a) $T=0~\text{K}$, (b) $50~\text{K}$, and (c) $100~\text{K}$.
        The free energy of ice VIII is set to zero as the reference.
        (d) The O-H bond lengths $d_{\text{OH}}$ plotted against the O-O distances $d_{\text{OO}}$.
        (e) The order parameter $\Delta = |d'_\text{OH} - d_\text{OO}/2|$ for ice VIII, where $d'_\text{OH}$ is the projection of $d_{\text{OH}}$ onto the O-O direction.
        (f) The equation of state.
        Experimental results were obtained at room temperature~\cite{wolanin1997equation,sugimura2008compression,asahara2010thermoelastic}, while the NCT results were calculated at $100~\text{K}$. 
        The inset highlights the transition pressure. 
        }
	\label{fig2}
\end{figure}

\textit{Neural canonical transformations of high-pressure ice}.--
The phase transition from ice VIII/VII to ice X is primarily driven by the symmetrization of hydrogen bonds, transforming ice from a molecular to an atomic crystal~\cite{hirsch1986effect,lee1992ab,lee1993ab,goncharov1996compression,aoki1996infrared,benoit1998tunnelling,bernasconi1998abinitio,loubeyre1999modulated,pruzan2003phase,caracas2008dynamical,lin2011correlated,meier2018observation,li2024phase,komatsu2024hydrogen,bronstein2014quantum,kuwahata2024nuclear, cherubini2024quantum}.  
Due to the highly localized nature of this transition, finite-size effects are negligible for nuclei, as confirmed by studies showing no significant differences between supercells containing $16$, $64$, or $128$ $\text{H}_2\text{O}$~\cite{kang2013quantum, kuwahata2024nuclear, cherubini2024quantum}. 
Therefore, our analysis will focus on a cell with $16$ $\text{H}_2\text{O}$.
Unlike the proton order-disorder transition between ice VIII and VII, driven by thermal fluctuations, the transition to ice X is governed exclusively by nuclear quantum effects. 
Thus, we only consider only proton-ordered configurations.


We performed variational free energy calculations for ice VIII and X at $T=0, 50, 100~\text{K}$ across a range of densities using the NCT method.
The initial configurations were based on ideal crystalline structures, which were subsequently optimized.
In ice X, hydrogen atoms are symmetrically positioned between oxygen atoms, while in ice VIII, they are asymmetrical.
Figures~\ref{fig2}~(a)-(c) show the relative free energies, with the free energy of ice VIII set as the reference.
At low pressures, which correspond to larger volumes, the ice VIII structure is more stable. However, ice X exhibits lower free energies at higher pressures. 
The phase transition pressure, determined from the intersection of free energy curves, is nearly temperature-independent and occurs at $60.6~\text{GPa}, $which is consistent with the experiments~\cite{aoki1996infrared,goncharov1996compression,struzhkin1997cascading,song1999infrared}.

The discontinuity in the second derivative of free energy indicates that this is a first-order transition. This observation is consistent with SSCHA calculations~\cite{ackland2025distinction} and a recent free energy model study~\cite{cherubini2024quantum}, but it diverges from earlier findings~\cite{pruzan2003phase}.
This feature also accounts for the negligible finite-size effects observed in the VIII-X transition. It is worth noting that this transition is distinct from the order-disorder transition described in Ref.~\cite{zhu2025quantum}, where achieving convergence requires extremely large supercells.
While PIMD encounters ergodicity challenges in first-order transitions, the NCT method is still robust.

To further quantify these phases, 
Fig.~\ref{fig2}~(d) illustrates the relationship of $d_\text{OH}$ and $d_\text{OO}$, with definitions provided in Fig.~\ref{fig1}~(a).
The progression from right to left represents the compression, with data across different temperatures collapsing into a single trend.
Initially, as pressure increases, $d_\text{OH}$ slightly increases while $d_\text{OO}$ decreases, which is characteristic of hydrogen-bonded systems. 
The much stronger covalent bond ($\text{O}-\text{H}$) resists compression, while the weaker hydrogen bond ($\text{O}\cdots\text{H}$) compresses more readily.
As the hydrogen atom shifts toward the midpoint between the two oxygen atoms, this difference leads to an anomalous elongation of the covalent bond~\cite{li2011quantum}.
A transition occurs at $d_\text{OO} \approx 2.38~\AA$.
Beyond this point, $d_\text{OH}$ decreases with $d_\text{OO}$ and approaches $d_\text{OO}/2$. 
This change indicates that the hydrogen atoms transition from being localized in single potential wells to being symmetrically distributed between two oxygen atoms.
As a result, the molecular structure of ice is disrupted, leading to the formation of an atomic crystal.


It is noteworthy that $d_\text{OO}$ is not exactly twice $d_\text{OH}$ in the X phase, consistent with observations in Ref.~\cite{kuwahata2024nuclear}.
The deviation arises from fluctuations of hydrogen atoms perpendicular to the O-O direction.
To account for this, we project the hydrogen position to the O-O direction and denote the projected distance as $d_\text{OH}'$.
The order parameter is then defined as $\Delta=|d_\text{OH}' - d_\text{OO}/2|$, 
which approaches zero for symmetric hydrogen-bonded structures,
as shown in Fig.~\ref{fig2}~(e).

We also determined the pressure-volume relationship, i.e., the equation of state (EOS), as illustrated in Fig.~\ref{fig2}~(f). 
The results reveal a discontinuity at $60.6~\text{GPa}$, highlighting the transition to the ice X phase.
At high pressures, our calculations align closely with experimental data~\cite{wolanin1997equation,sugimura2008compression,asahara2010thermoelastic}. 
Below $60~\text{GPa}$, values from different experiments exhibit a spread of approximately $2$ to $5~\text{GPa}$ at each volume, and the NCT results fall within this range.

\begin{figure}[htbp]
	\centering
    \hspace{-4.0mm}
	\includegraphics[width=0.50\textwidth]{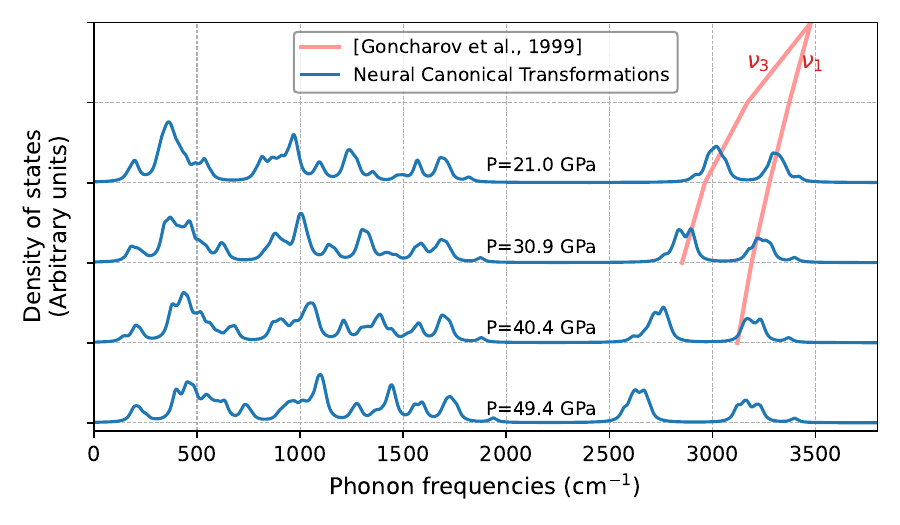}
	\caption{
    Phonon density of states for ice VIII under pressure.
    Anharmonic phonon frequencies from Neural Canonical Transformations are derived from single-phonon excitation energies. 
    Experimental data are taken from Refs.~\cite{goncharov1999raman}, with molecular O-H vibrational modes denoted as $\nu_1$ and $\nu_3$.
    }
	\label{fig3}
\end{figure}



One of the key advantages of the NCT method is its ability to calculate phonon spectra beyond the harmonic approximation. 
The anharmonic phonon frequencies are determined by the energy difference between the ground state and the single-phonon excited state, $\omega_k^{\text{anh}} = E_{n_k=1} - E_{n_k=0}$.
Here, $E_{\bm{n}} = \mathbb{E}[H_{\text{vib}}\Psi_{\bm{n}}/\Psi_{\bm{n}}]$ represents the energy expectation value, following Ref.~\cite{zhang2024lithium}, where ${\mathbb{E}}$ denotes the statistical expectations. 

As shown in Fig.~\ref{fig3},
for the ice VIII structure at $21.0~\text{GPa}$, high-frequency peaks around $3000~\text{cm}^{-1}$ and $3300~\text{cm}^{-1}$ correspond to the symmetric and asymmetric O-H stretching modes ($\nu_1$, $\nu_3$) of the water molecule, which are in excellent agreement with experimental observations~\cite{goncharov1999raman}.
Experimentally, ice VIII exhibits anomalous softening of these stretching modes under pressure. At lower pressures, the $\nu_1$ and $\nu_3$ are nearly degenerate. As pressure increases, the frequency of both modes decreases, with $\nu_3$ softening more rapidly than $\nu_1$. 
This behavior contrasts with typical systems where pressure typically shortens chemical bonds and hardens phonon modes.
The redshift of the stretching frequency ($\nu_1$, $\nu_3$) is a widely recognized indicator of hydrogen bond systems~\cite{xantheas1993ab,li2011quantum}.
It is attributed to an increase in the covalent bond length and a weakening of covalent interactions as pressure increases.
In the low-frequency region, the NCT calculations identify a crystal translational mode ($\nu_{\text{T}}$) and rotation mode ($\nu_{\text{R}}$) around $500~\text{cm}^{-1}$ and $1000~\text{cm}^{-1}$, respectively, which shifts to higher frequencies with increasing pressure.

\textit{Gaussian and non-Gaussian-type wavefunctions}.--
Building on our prior demonstration that non-Gaussian flow models achieve lower ground-state energies than Gaussian wavefunctions for a one-dimensional strongly anharmonic double-well potential~\cite{zhang2024lithium}, we extend this comparison to quantum solids. 
For clarity, our analysis focuses exclusively on zero-temperature conditions in the phonon coordinates. 

Within the NCT framework, the variational ground-state wavefunction is expressed as a product of Gaussian distributions and the square root of the Jacobian determinant~\cite{zhang2024neural, zhang2024lithium, supp}:
$\Psi(\bm{q})= 
\prod_{k}
e^{-\frac{1}{2} \omega_k \xi_k^2}
\sqrt{
\left| 
\text{det} (\partial \bm{\xi} / \partial \bm{q})
\right|
}$,
where $\omega_k$ is the harmonic frequency of the $k$ th phonon, calculated from the dynamical matrix.
Here, $\bm{q}=\{q_k\}$ are phonon coordinates, and $\bm{\xi}=\{\xi_k\}$ denotes the quasiphonon coordinates.
For non-Gaussian calculation,
the transformation $\bm{\xi}=f_{\bm{\theta}}(\bm{q})$ is parameterized by a real-valued non-volume preserving network~\cite{dinh2016density}, which effectively captures anharmonicity and phonon interactions.
In contrast, the self-consistent Gaussian approximation reduces to mode-wise linear transformations:
$\xi_k= a_k q_k + b_k$, where $a_k$ and $b_k$ are trainable parameters of each mode, while neglecting phonon interactions.


\begin{figure}[htbp]
	\centering
    \hspace{-4.0mm}
	\includegraphics[width=0.50\textwidth]{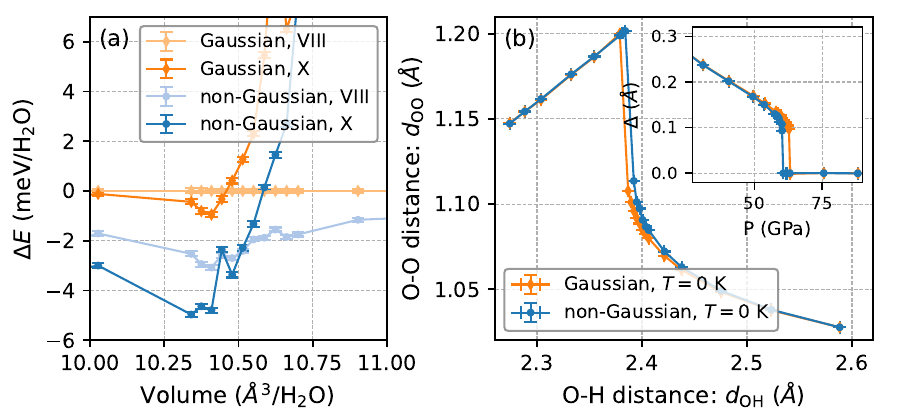}
	\caption{
    Numerical results of NCT with non-Gaussian and Gaussian normalizing flow model.
    (a) Energy differences between ice VIII and X structures at zero temperature, with ice VIII (Gaussian) set to zero as a reference.
    (b) O-H bond lengths $d_{\text{OH}}$ versus O-O distances $d_{\text{OO}}$, with the inset showing the order parameter $\Delta = |d'_{OH} - d_{OO}/2|$.
    }
	\label{fig4} 
\end{figure}

As shown in Fig.~\ref{fig4}~(a),
the non-Gaussian wavefunction consistently enables NCT to achieve lower energies than the Gaussian approximation.
For ice VIII, the non-Gaussian calculations yield energies approximately $2~\text{meV}$ lower than the Gaussian approximation.
This difference becomes more pronounced in ice X, where non-Gaussian energies are $3$ to $4~\text{meV}$ lower.
The larger discrepancy in ice X arises from its stronger anharmonicity, where hydrogen atoms are delocalized over a double-well potential rather than confined to a single minimum. 
The Gaussian approximation cannot effectively capture these delocalized quantum wavefunctions, resulting in higher energy predictions.
Besides overestimating energy, the Gaussian approximation also introduces minor differences in the bond distances and order parameters, as shown in Fig.~\ref{fig4}~(b).
Additionally, analysis of the order parameter $\Delta$ reveals a reduction of approximately $2~\text{GPa}$ in the VIII-X transition pressure compared to the non-Gaussian NCT calculations.


\textit{Conclusions and discussions}.--
This study investigates high-pressure ice phases, providing insights into hydrogen-containing compounds under extreme conditions, particularly the symmetrization of hydrogen bonds. We analyze various factors that may influence the VIII-X phase transition.
Classical simulations predict a transition pressure exceeding $100~\text{GPa}$~\cite{kuwahata2024nuclear, cherubini2024quantum}, 
far above the experimental results around $60~\text{GPa}$~\cite{aoki1996infrared,goncharov1996compression,struzhkin1997cascading,song1999infrared}.
This discrepancy highlights the essential role of nuclear quantum effects, which enable hydrogen quantum tunneling in the double-well potential.
In electronic structure calculations, 
we provide the first conclusive demonstration that SCAN achieves accuracy comparable to QMC in high-pressure ice.
The PBE functional underestimates the double-well potential barrier relative to SCAN, resulting in an over $10~\text{GPa}$ underestimation of the transition pressure.
Incorporating quantum anharmonic effects through NCT calculations, we find that the transition pressure to ice X is temperature-independent, consistent with experimental results. This observation further indicates that the transition is entirely a quantum phase transition.
Calculations using Gaussian-type wavefunctions reveal a $2~\text{GPa}$ increase in the predicted transition pressure compared to non-Gaussian calculations performed with normalizing flow models.

This Letter also showcases the successful application of the NCT method to high-pressure ice with hydrogen-bond symmetrization and phonon softening in the hydrogen-oxygen stretching modes. 
These results underscore its ability to study anharmonic quantum solids.
Despite its strengths, there is still room for improvement, particularly in the variational wavefunctions.
Although the non-Gaussian normalizing flow model outperforms the Gaussian ansatz, its fixed topological structure for basis states limits the expressiveness of the wavefunction.
Integrating vibrational self-consistent field theory~\cite{bowman1978self, bowman1986self, gerber1979semiclassical, monserrat2013anharmonic, monserrat2014electron, hutcheon2019structure, kapil2019assessment} into NCT calculations, could enable wavefunction mixing with diverse nodal structures.
Additionally, current calculations struggle to resolve Fermi resonance, which requires the mixing of different modes. Employing vibrational configuration interaction~\cite{christoffel1982investigations} wavefunctions could effectively address this limitation.
Finally, the ability of NCT to compute phonon spectra could be extended to electron-phonon coupling calculations, a critical factor for understanding superconductivity, particularly in hydride solids~\cite{ion2013first, ion2014anharmonic, errea2016quantum, errea2020quantum}.

\textit{Acknowledgements}.--
We are grateful for the useful discussions with Han Wang and Zhendong Cao. The DFT calculations have been done on the supercomputing system in the Huairou Materials Genome Platform. This work is supported by the National Natural Science Foundation of China under Grants No. 92270107, No. T2225018, No. 12188101, No. T2121001 and the Strategic Priority Research Program of the Chinese Academy of Sciences under Grants No. XDB0500200, and the National Key Projects for Research and Development of China Grants No. 2021YFA1400400. 


\bibliographystyle{apsrev4-1}
\bibliography{mybibtex}

\clearpage

\appendix
\begin{widetext}
\begin{center}
    {\large \textbf{Supplemental Material: Quantum Anharmonic Effects in Hydrogen-Bond Symmetrization of High-Pressure Ice
    }}
\end{center}

\setcounter{table}{0}
\setcounter{figure}{0}
\setcounter{equation}{0}
\setcounter{section}{0}
\renewcommand{\theequation}{S\arabic{equation}}
\renewcommand{\thefigure}{S\arabic{figure}}
\renewcommand{\thetable}{S\arabic{table}}
\renewcommand{\thesection}{S\Roman{section}}


\section{A. Electronic structure calculations}

We conducted electronic structure calculations using the Vienna Ab initio Simulation Package (VASP)~\cite{vaspref} with projector-augmented wave pseudopotentials. 
To investigate the impact of exchange-correlation functionals, calculations were conducted using
the Perdew-Burke-Ernzerhof (PBE)~\cite{perdew1996generalized} and the strongly constrained and appropriately normed semilocal density (SCAN)~\cite{sun2015strongly} functionals under identical computational settings. 
The plane-wave cutoff energy was set to $1500~\text{eV}$, and the $k$-point spacing was set as $0.2~\text{\AA}^{-1}$ to ensure convergence.

The quantum Monte Carlo simulations are performed using the DeepSolid code~\cite{li2022deepsolid}, which employs a periodic neural network as the wavefunction ansatz for solid systems, offering exceptional accuracy.
The neural network architecture consists of $4$ layers,
with the single-electron layer dimension set to $256$ and the two-electron layer dimension set to $32$.
All other parameters are set to the default values specified in the Ref.~\cite{li2022deepsolid}.

\section{B. Machine learning potential}

\begin{figure*}[htbp]
	\centering
	\includegraphics[width=0.7\textwidth]{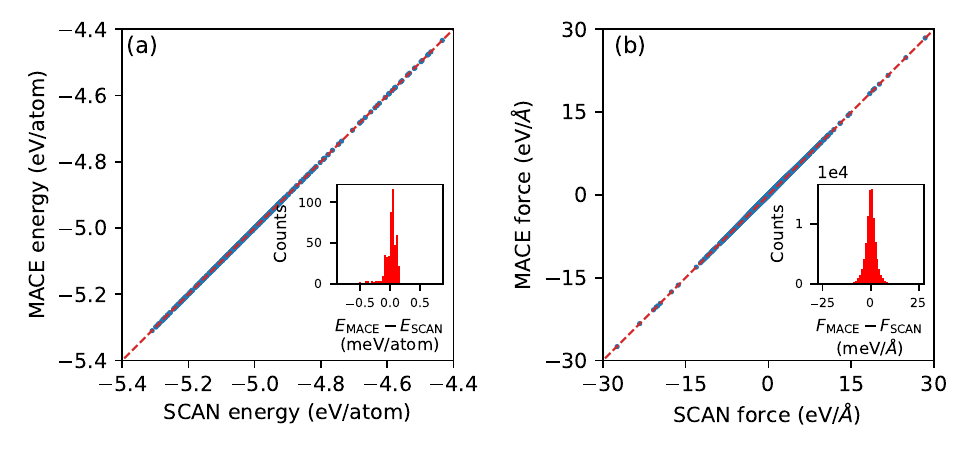}
    \caption{Comparison between MACE predictions and density functional theory calculations with SCAN functional on the test dataset: (a) energy, (b) force.}
	\label{Fig:S1}
\end{figure*}

In this work,
a total of $9081$ configurations were calculated through VASP calculations~\cite{vaspref} using the SCAN functional~\cite{sun2015strongly,sun2015scan}, covering both ice VIII and X structures with cell sizes of $16$ to $128$ $\text{H}_2\text{O}$ and pressures between $20$ and $200~\text{GPa}$.
To achieve a balance between accuracy and efficiency, we employed the high-order equivariant message-passing neural networks, known as multi-atomic cluster expansion (MACE)~\cite{batatia2022mace, batatia2022design}, to fit both energy and force data.
The dataset comprises $8627$ configurations for training and $478$ for testing.
With a $4.0~\AA$ radial cutoff and a two-layer structure, the model's receptive field extended to $8.0~\AA$.
As shown in Fig.~\ref{Fig:S1},
the MACE model achieved excellent performance, achieving a root mean square error of $0.1~\text{meV}/\text{atom}$ for energy and $3~\text{meV}/\AA$ for forces on the test set.

\section{C. Neural canonical transformation}
The Hamiltonian for nuclear vibrations of water ice can be written as:
$
 H_{\text{vib}} = - \sum_i \frac{1}{2M_i} \nabla_i^2 + V(\bm{R}),
 $
where $M_i$ are the nuclear masses, with hydrogen and oxygen set to $1.008~\text{amu}$ and $15.995~\text{amu}$, respectively.
The PES is $V(\bm{R})$ computed using MACE.
By diagonalizing the dynamical matrix~\cite{martin2020electronic}, 
the Hamiltonian is reformulated in the harmonic frequencies and phonon coordinates:
\begin{equation}
H_{\text{vib}} 
= \frac{1}{2} \sum_{k=1}^D \left(
    - \frac{\partial^2 }{ \partial q_k^2} 
    + \omega_k^2 q_k^2
    \right)
+ V_{\text{anh}}(\bm{q}),
\end{equation}
where $D$ is the number of vibrational modes in a supercell, $\omega_k$ is the harmonic frequency of the $k$ th phonon ($k=1,2,...,D$), and $V_{\text{anh}}$ represents the anharmonic contributions in PES.
In the phonon representation $\bm{q}=(q_1, q_2, ..., q_D)$, the vibrational modes of the crystal are effectively decoupled from those of the molecules.
Notably, the stretching mode of the O-H bond is distinctly separated.

The solution at finite temperature $T$ can be obtained using the NCT method, a variational free energy approach detailed in our previous work~\cite{zhang2024neural,zhang2024lithium}.
NCT employs Monte Carlo sampling to provide an unbiased estimate of the anharmonic free energy, formulated as nested thermal and quantum expectations:
\begin{equation}
F = \underset{\bm{n} \sim p_{\bm{n}}}
    {\mathbb{E}} 
    \left[
    k_B T \ln p_{\bm{n}} + 
    \underset{\bm{q} \sim |\Psi_{\bm{n}}(\bm{q})|^2}
    {\mathbb{E}} \left[
        \frac{H_{\text{vib}} \Psi_{\bm{n}}(\bm{q})}{\Psi_{\bm{n}}(\bm{q})}
    \right]
\right],
\label{eq:helmholtz}
\end{equation}
where $\bm{n}=(n_1, n_2, \ldots, n_D)$ indexes the phonon energy levels,
and ${\mathbb{E}}$ denotes the statistical expectations.
The variational parameters controlling the energy occupation probabilities are denoted by $\bm{\mu}$, with $p_{\bm{n}}=p_{\bm{n}}({\bm{\mu}})$.
A normalizing flow~\cite{dinh2014nice, dinh2016density, papamakarios2019neural, papamakarios2021normalizing, saleh2023computing} parameterizes the wavefunctions $\Psi_{\bm{n}}(\bm{q})=\Psi_{\bm{n}}(\bm{\theta},\bm{q})$ with model parameters $\bm{\theta}$.
Both $\bm{\mu}$ and $\bm{\theta}$ are optimized via stochastic gradient descent~\cite{kingma2014adam}, with $F$ serving as the loss function.

The central idea of NCT is to construct a reversible transformation between simple basis and complex anharmonic phonon wavefunctions.
Specifically, the orthogonal variational wavefunctions for all energy levels are given by~\cite{zhang2024neural,zhang2024lithium}:
\begin{equation}
\Psi_{\bm{n}}(\bm{q})
= \Phi_{\bm{n}}\left(f_{\bm{\theta}}(\bm{q})\right) \left| \mathrm{det}
\left(
\frac{\partial f_{\bm{\theta}}(\bm{q})}{\partial \bm{q}}
\right)
\right|^{1/2},
\label{eq:flow}
\end{equation}
where the basis states $\Phi_{\bm{n}}$ are chosen as wavefunctions of a $D$-dimensional harmonic oscillator with frequencies $\omega_k$, and the Jacobian determinant incorporates phonon interactions and anharmonic effects.
Importantly, the transformation is achieved by a coordinate mapping between the phonon coordinates $\bm{q}$ and the quasiphonon latent space $\bm{\xi}$.
The bijection $\bm{\xi}=f_{\bm{\theta}}(\bm{q})$ is smooth, parameterized by $\bm{\theta}$, and can be realized using real-valued non-volume-preserving networks~\cite{dinh2016density}.

\section{D. Additional characterization of the double-well potential with different functionals}
\begin{figure*}[htbp]
	\centering
	\includegraphics[width=0.7\textwidth]{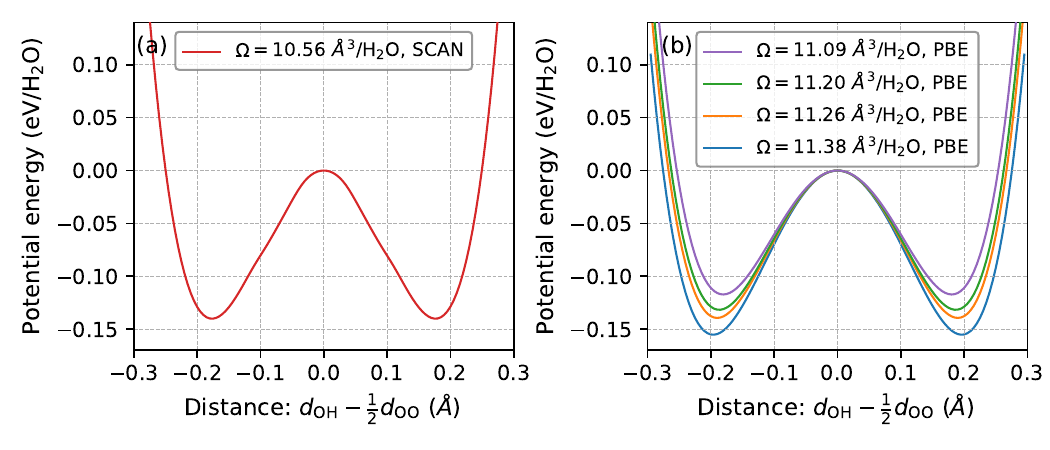}
    \caption{The potential energy surface is calculated using (a) SCAN functional, (b) PBE functional.
    All hydrogen atoms are moved along the O-O direction, while the oxygen atoms remain fixed.
    }
	\label{Fig:S0}
\end{figure*}

As shown in Fig.~\ref{Fig:S0}, our fixed-volume calculations reveal that the SCAN functional predicts a barrier height of $140~\text{meV}$ at $\Omega=10.56\AA^3/\text{H}_2\text{O}$ ($60~\text{GPa}$), which allows hydrogen tunneling to occur in the high-pressure ice.
In contrast, the PBE functional requires $\Omega = 11.26~\text{\AA}^3/\text{H}_2\text{O}$ ($47~\text{GPa}$) to achieve the same barrier height. 
The pressures corresponding to these volumes are estimated using the equation of state obtained from NCT calculations with SCAN functional.

\end{widetext}

\end{document}